# Superconductivity in a misfit phase combining a topological crystalline insulator and a layered transition metal dichalcogenide

Huixia Luo[1,*], Kai Yan[2], Weiwei Xie[1], Brendan F. Phelan[1], and R. J. Cava[1,*]

[1]Department of Chemistry, Princeton University, Princeton, NJ 08544, USA

[2]School of Engineering, Brown University, 182 Hope Street, Providence, RI 02912, USA

**Abstract**

We report the characterization of the misfit compound $(Pb_{1-x}Sn_xSe)_{1.16}(TiSe_2)_2$ for $0 \leq x \leq 0.6$, in which a [100] rocksalt bilayer of $Pb_{1-x}Sn_xSe$, which is a topological crystalline insulator in bulk form, alternates with a double layer of the normally non-superconducting layered transition metal dichalcogenide $TiSe_2$. The x dependence of $T_c$ displays a weak dome-like shape with a maximum $T_c$ of 4.5 K at *x* = 0.2; there is only a subtle change in Tc corresponding to the expected trivial to topological transition in the $Pb_{1-x}Sn_xSe$ bi-layer. We present a more detailed characterization of the superconductor at x = 0.4, for which the bulk $Pb_{1-x}Sn_xSe$ phase is in the topological crystalline insulator regime. For this material, the Sommerfeld parameter $\gamma = 11.06$ mJ mol$^{-1}$ K$^{-2}$, the Debye temperature $\Theta_D = 161$ K, the normalized specific heat jump value $\Delta C/\gamma T_c = 1.38$ and the electron-phonon constant value $\lambda_{ep} = 0.72$, suggesting that $(Pb_{0.6}Sn_{0.4}Se)_{1.16}(TiSe_2)_2$ is a BCS-type weak coupling superconductor. This material is of interest for probing the interaction of superconductivity with the surface states of a topological crystalline insulator.

1. **Introduction**

A broad family of layered ternary chalcogenides, the so-called misfit compounds, has recently been reported. They are generally described as $[(MX)_{1+x}]_m(TX_2)_n$, where M = Sn, Pb, Sb, Bi or Ln (lanthanide); T = Ti, V, Cr, Nb, Mo, Ta, or W; X = S, Se or Te; 0.08 < x < 0.28; and $m$ and $n$ are integers indicative of the number of MX rocksalt double layers stacked in an alternating fashion with $TX_2$ dichalcogenide layers ($n$ and $m$ =1, 2, 3, 4) [1-8] The MX and $TX_2$ layers have different symmetry and periodicity, matching in size in one crystallographic in-plane direction but not matching (i.e. misfitting) in the second in-plane direction, yielding, even though the materials are fully crystallographically ordered, the oddly non-stoichiometric formulas. The individual rocksalt and dichalcogenide layers in the misfit compounds are not significantly structurally distorted from those of the constituent simple MX and $TX_2$ materials, meaning that the electronic structures of the constituent layers are expected to be equivalent to those of the individual bulk MX and $TX_2$ phases, though charge may be transferred from one layer to the other.

The wide variations in M, T, X, $m$ and $n$ allowed in the family lead to many different physical properties. [9-17] The current case involves selenide misfits based on the stacking of [100] PbSe double layers and [001] $TiSe_2$ layers. PbSe is a trivial semiconductor (i.e the sequence of the electronic bands near the Fermi energy is as expected in a simple picture) and has a direct band gap of 0.27 eV at room temperature. [18,19] Recently it has been shown, however, that the $Pb_{1-x}Sn_xSe$ rocksalt compound undergoes an inversion of the electronic band energy sequence at x = 0.23 and becomes a topological crystalline insulator, with, correspondingly, protected topological surface states. [19] When PbSe is stacked in a misfit compound with $NbSe_2$, the intrinsic $T_c$ of $NbSe_2$ (7K) is degraded and superconductivity results for $(PbSe)_{1.14}(NbSe_2)_n$ for n = 2 and 3 at 3.4 and 4.8 K respectively, with no reported superconductivity for n = 1 down to 2 K.[20] When stacked with non-superconducting $1T$-$TiSe_2$,[21-24] on the other hand, the resulting misfit compound $(PbSe)_{1.16}(TiSe_2)_2$ is reported to be a superconductor with $T_c$ = 2.3 K.[25] Motivated by these observations, here we report the superconductivity that results when Sn partially substitutes for Pb in the $(Pb_{1-x}Sn_xSe)_{1.16}(TiSe_2)_2$ misfit compound, in order to determine whether there is a discontinuity in the superconducting $T_c$ when the trivial to topological transition occurs in bulk $Pb_{1-x}Sn_xSe$, which occurs as bilayers in the misfit compounds. We observe only a subtle change in $T_c$ vs. x at that composition.

2. **Experimental**

$(Pb_{1-x}Sn_xSe_2)_{1.16}(TiSe_2)_2$ single crystals and polycrystals were grown in three steps. First, mixtures of high-purity fine powders of Pb (99.9%), Sn (99.5%), Ti (99.9%) and Se (99.999%) in the appropriate stoichiometric ratios were mixed and heated in sealed evacuated silica tubes at a rate of 1 °C/min to 700 °C and held there for 48 h. Subsequently, the as-prepared powders were reground and heated at a rate of 3 °C/min to 900 °C and held there for 16 h. Finally, larger and smaller crystals from the as-prepared powders were grown in the third step by the chemical vapor transport (CVT) method, using $SeCl_4$ as a transport agent. 50 mg of the as-prepared powders of $(Pb_{1-x}Sn_xSe_2)_{1.16}(TiSe_2)_2$ were mixed with 35 mg $SeCl_4$, sealed in evacuated silica tubes and heated for one week in a two-zone furnace, where the temperature of source and growth zones were fixed at 750 °C and 680 °C, respectively. After one week, polycrystalline samples and some shiny, plate-like grey single crystals of $(Pb_{1-x}Sn_xSe_2)_{1.16}(TiSe_2)_2$ were found at the cold end. Property measurements were performed on the crystals and polycrystals (collections of small single crystals) from the cold end.

The identity and phase purity of the materials studied were determined by powder X-ray diffraction (PXRD) on polycrystals and crystal plates using a Bruker D8 ECO diffractometer with Cu Kα radiation and a Lynxeye detector. To determine the phase purity, LeBail fits were performed on the powder diffraction data through the use of the FULLPROF diffraction suite using Thompson-Cox-Hastings pseudo-Voigt peak shapes.[26] Single crystals selected from partially crushed crystalline samples were studied on a Bruker D8 ECO single crystal diffractometer with Cu Kα radiation to fully verify that the materials matched the misfit structure previously reported for the Pb-containing end member. The compositions of the materials were determined by employing Energy dispersive X-ray fluorescence spectroscopy (EDX) on the crystals using a Quanta 200 FEG ESEM electron microscope operated at 20 kV. Measurements of the temperature dependence of the electrical resistivity and specific heat were performed in a Quantum Design Physical Property Measurement System (PPMS). Zero-field cooled (ZFC) and field cooled (FC) magnetic susceptibilities were measured in a field of 10 Oe using a Quantum Design superconducting quantum interference device (SQUID) magnetometer.

3.  **Results and Discussion**

Figure 1 shows the comparison of the Pb/Ti and Sn/Ti ratios obtained from the EDX measurements and the starting material compositions. The Pb/Ti and Sn/Ti ratios obtained from the EDX results were found to be within experimental error of the ratios present in the starting materials, indicating that the compositions of the $(Pb_{1-x}Sn_xSe)_{1.16}(TiSe_2)_2$ crystals and polycrystals grown between x = 0 and x = 0.6 are within error of the nominal compositions. Single phase crystals of the $(Pb_{1-x}Sn_xSe)_{1.16}(TiSe_2)_2$ misfit phase could only be obtained by our method up to x = 0.6. For higher x, the $((Pb,Sn)Se)_{1.16}(TiSe_2)_2$ misfit crystals were multiple-phase. Pure crystals of $(SnSe)_{1.16}(TiSe_2)_2$ were also obtained, but they were not superconducting above 1.8 K, and are not the subject of this study.

A representative room temperature X-ray diffraction pattern (PXRD), obtained from a crystal plate of $(Pb_{0.6}Sn_{0.4}Se)_{1.16}(TiSe_2)_2$ (example shown in the inset), looking at diffraction from the (00l) planes, is shown in Figure 2. The pattern is very similar to that reported previously for the known $TiSe_2$ double layer misfit $(PbSe)_{1.16}(TiSe_2)_2$ [25] confirming the misfit character of the materials characterized here. Figures 3 a and b show schematics of the misfit crystal structure of the $(Pb_{1-x}Sn_xSe)_{1.16}(TiSe_2)_2$, (using the example of x = 0.4) viewed in different directions. The figures highlight the basic structure as an alternating stacking of $(Pb_{1-x}Sn_x)Se$ rocksalt bi-layers with two 1T-like $TiSe_2$ layers, and also the incommensurate nature of their in-plane matching.

**Figure 4** shows the systematic change in the transport properties of $(Pb_{1-x}Sn_xSe)_{1.16}(TiSe_2)_2$ on increasing $x$. **Figure 4a** shows the temperature dependence of the resistivity ratio ($\rho/\rho_{300K}$) for polycrystalline $(Pb_{1-x}Sn_xSe_2)_{1.16}(TiSe_2)_2$ (0.0 ≤ x ≤ 0.6). The inset to the figure enlarges the resistivity behavior in the low temperature region (2 – 5.5 K); showing the superconducting transition. **Figure 4b** shows the d$\rho$/dT for $(Pb_{1-x}Sn_xSe)_{1.16}(TiSe_2)_2$ (0.0 ≤ x ≤ 0.6) in the low temperature region (2 – 5.5 K), further showing the superconducting transition. At low temperatures, a clear, sharp ($\Delta T_c$ < 0.5 K) drop of $\rho$(T) is observed, signifying the onset of superconductivity. We find that the $T_c$ changes just slightly with the increase of doped Sn content x, displaying a very weak dome-shaped peak at intermediate compositions. We note that the Tc observed here for x = 0 is higher than the one previously reported for that material; the reason for the improved Tc is not currently known.

The characterization of the superconducting properties of single crystal $(Pb_{0.6}Sn_{0.4}Se)_{1.16}(TiSe_2)_2$ by specific heat is shown in **Figure 5**. The main panels of **Figure 5a** show the temperature dependence of the specific heat ($C_p$/T versus $T^2$) under zero-field and under 5 Tesla field for $(Pb_{0.6}Sn_{0.4}Se)_{1.16}(TiSe_2)_2$. The normal state specific heat at low temperatures (but above $T_c$) obeys the relation of $C_p = \gamma T + \beta T^3$, where $\gamma$ and $\beta$ characterize the electronic and phonon contributions, respectively, the latter of which is a measure of the Debye Temperature ($\theta_D$). By fitting the data in the temperature range of 2 - 5 K, we obtain the electronic specific heat coefficient $\gamma$ = 11.06 mJ·mol$^{-1}$·K$^{-2}$ for $(Pb_{0.6}Sn_{0.4}Se)_{1.16}(TiSe_2)_2$. Per TiSe$_2$ layer, this number (11.06/2 ~ 5.5 mJ·mol$^{-1}$·K$^{-2}$) is significantly larger than that found for other superconductors based on TiSe$_2$ (i.e. 4.3 mJ·mol$^{-1}$·K$^{-2}$ for Cu$_{0.08}$TiSe$_2$ [27] and 2 mJ·mol$^{-1}$·K$^{-2}$ for Ti$_{0.8}$Ta$_{0.2}$Se$_2$ [28]).

The superconducting transition temperature observed in the specific heat measurements for $(Pb_{0.6}Sn_{0.4}Se)_{1.16}(TiSe_2)_2$ is in excellent agreement with the $T_c$ determined in the $\rho$(T) measurements. From the inset in **Figure 5**, using the equal area construction method, we obtain $\Delta C/T_c$ = 11.28 mJ·mol$^{-1}$·K$^{-2}$ for $(Pb_{0.6}Sn_{0.4}Se)_{1.16}(TiSe_2)_2$. The normalized specific heat jump value $\Delta C/\gamma T_c$ is thus found to be 1.38 for $(Pb_{0.6}Sn_{0.4}Se)_{1.16}(TiSe_2)_2$, which confirms the bulk superconductivity. This value is smaller than that of the Bardeen-Cooper-Schrieffer (BCS) weak-coupling limit value (1.43), but is in a range typically observed in complex materials. Using the Debye temperature ($\theta_D$), the critical temperature $T_c$, and assuming that the electron-phonon coupling constant ($\lambda_{ep}$) can be calculated from the inverted McMillan formula:[29]

$$\lambda_{ep} = \frac{1.04 + \mu^* \ln\left(\frac{\theta_D}{1.45 T_C}\right)}{(1 - 0.62\mu^*)\ln\left(\frac{\theta_D}{1.45 T_C}\right) - 1.04},$$

the value of $\lambda_{ep}$ obtained is 0.72 for $(Pb_{0.6}Sn_{0.4}Se)_{1.16}(TiSe_2)_2$. This suggests weak coupling superconductivity. The density of states at the Fermi level ($N(E_F)$) can be calculated from the following equation:

$$N(E_F) = \frac{3}{\pi^2 k_B^2 (1 + \lambda_{ep})} \gamma,$$

by using the value of Sommerfeld parameter ($\gamma$) and the electron-phonon coupling ($\lambda_{ep}$). This yields $N(E_F)$ = 2.73 states/eV f.u. for $(Pb_{0.6}Sn_{0.4}Se)_{1.16}(TiSe_2)_2$.

Finally, the electronic phase diagram as a function of temperature and doping level for the $(Pb_{1-x}Sn_xSe)_{1.16}(TiSe_2)_2$ misfit phase series is summarized in **Figure 6**. It can be seen that in the $(Pb_{1-x}Sn_xSe)_{1.16}(TiSe_2)_2$ system, the x dependence of $T_c$ displays a dome-like shape that is broad in composition. However, $T_c$ changes only slightly with the increase content of doped Sn x, with a maximum $T_c$ of 4.5 K at $x$ = 0.2. If there is any change on crossing the trivial to topological composition regime in the $Pb_{1-x}Sn_xSe$ bilayer at x = 0.25 it is only a subtle change in the slope of Tc vs x near the top of the dome; the most significant observation in this regard being the fact that for x = 0.4, well into the topological regime of $Pb_{1-x}Sn_xSe$, the misfit material remains superconducting. For compositions between x = 0.6 and x=1.0 we could not obtain single phase materials, and thus no data is shown in the figure for higher x; single phase misfit material was obtained at x = 1 (i.e. $(SnSe)_{1.16}(TiSe_2)_2$), but we did not observe any superconducting transition down to 2 K.

## Conclusion

The misfit phase $(Pb_{1-x}Sn_xSe)_{1.16}(TiSe_2)_2$ ($0 \leq x \leq 0.6$) series, which combines layers of the rocksalt structure topological crystalline insulator $Pb_{1-x}Sn_xSe$ with the layers of the transition metal dichalcogenide $TiSe_2$, is reported, and the trends in superconductivity in the series characterized. The superconducting transition temperature shows a weak dome shape with varying x, with a maximum $T_c \approx 4.5$ K very close to the expected trivial to inverted transition in the band structure of the rocksalt layers. For the misfit superconductor $(Pb_{0.6}Sn_{0.4}Se)_{1.16}(TiSe_2)_2$, which is well within the topological composition regime of the rocksalt layers, the Sommerfeld parameter $\gamma = 11.06$ mJ mol$^{-1}$ K$^{-2}$, the Debye temperature $\Theta_D = 161$ K, the normalized specific heat jump value $\Delta C/\gamma T_c = 1.11$ and the electron-phonon constant value $\lambda_{ep} = 0.72$, suggesting that $(Pb_{0.6}Sn_{0.4}Se)_{1.16}(TiSe_2)_2$ is a BCS-type weak coupling superconductor. No superconducting transition is observed for the $(SnSe)_{1.16}(TiSe_2)_2$ misfit above 1.8 K. Why that SnSe-based misfit material is not superconducting at temperatures comparable to those exhibited by the Pb+Sn containing misfit phases is not currently known. Misfit phases in this system with compositions where the $Pb_{1-x}Sn_xSe$ rocksalt layer is in the topological crystalline insulator regime will be of interest for examining the relationship between TCIs and superconductivity.

## Acknowledgements

The materials synthesis and physical property characterization of this superconductor were supported by the Department of Energy, division of basic energy sciences, Grant DE-FG02-98ER45706. The authors acknowledge discussions with Ali Yazdani.

**References**

1. R. L. Withers, L. A. Bursil, *Philos. Mag.*, 1981, **B43 (4),** 635.
2. G. A. Wiegers and W. Y. Zhou, *Mat. Res. Bull.,* 1991, **26**, 879.
3. W. Y. Zhou, A. Meetsma, J. L. de Boer, and G. A. Wiegers, *Mat. Res. Bull.,* 1992, **27**, 563.
4. R. Atkins, S. Disch, Z. Jones, I. Haeusler, C. Grosse, S. F. Fischer, W. Neumann, P. Zschack, and D. C. Johnson, *J. Solid State Chem*., 2013,.**202**, 128.
5. F. R. Harris, S. Standridge, and D. C. Johnson, *J. Am. Chem. Soc*., 2005, **127**, 7843.
6. N. S Gunning, J. Feser, M. Falmbigl, M. Beekman, D. G Cahill, and D. C Johnson, *Semicond. Sci. Technol.*, 2014, **29(12)**, 124007.
7. T. Kondo, K. Suzuki and T. Enoki, *Solid State Com*, 1992, **84(10)**, 999.
8. G. A. Wiegers, A. Meesma, R. J. Haange, S. V. Smaalen and J. L. De Boer, *Acta. Cryst.*, 1990, **B46**, 324.
9. Y. Gotoh, M. Goto, K. Kawaguchi, Y. Oosawa and M. Onoda, *Mat. Res. Bull.,* 1990, **25**, 307.
10. M. Onoda, K. Kato, Y. Gotoh and Y. Oosawa, *Acta Cryst.,* 1990, **B46**, 487.
11. Y. Gotoh, J. Akimoto, M. Sakurai, Y. Kiyozumi, K. Suzuki and Y. Oosawa, Chem. Lett.,1990, **19(11)**, 2057.
12. Y. Gotoh, J. Akimoto, Y. Oosawa, M. Onoda, *Jpn. J. Appl. Phys.*, 1995, **34**, 662.
13. N. T. Nguyen, B. Howe, J. R. Hash, N. Liebrecht, P. Zschack, D. C. Johnson, *Chem. Mater*. 2007, **19**, 1923.
14. Y.Oosawa, Y Gotoh, J. Akimoto, M. Sohma, T., Tsunoda, H. Hayakawa and M. Onoda, *Solid State Ion.*, 1994, **67**,287.
15. D. B. Moore, M. Beekman, S. Disch, and D. C. Johnson, *Angew. Chem. Int. Ed*., 2014, *53*, 5672.
16. F. R. Harris, S. Standridge, and D. C. Johnson, *J. Am. Chem. Soc*., 2005, **127**, 7843.
17. K. Takita and K. Masuda, *J. Low Temp. Phys.*, 1985, **58(1),** 127.

18. P. Dziawa, B. J. Kowalski, K. Dybko, R. Buczko, A. Szczerbakow, M. Szot, E. Łusakowska, T. Balasubramanian, B. M. Wojek, M. H. Berntsen, O. Tjernberg, T. Story, *Nat. Mater.,* 2012, **11**, 1023.
19. W. W. Yu, J. C. Falkner, B. S. Shih, and V. L. Colvin, *Chem. Mater.,* 2004, **16**, 3318.
20. Y. Oosawa, Y. Gotoh, J. Akimoto, T. Tsunoda, M. Sohma, M. Onoda, *Jpn. J. Appl. Phys.* 1992, **31**, L1096.
21. F. J. Di Salvo, D. E. Moncton, J. V. Waszczak, *Phys. Rev. B,* 1976, **14**, 4321.
22. N. G. Stoffel, S. D. Kevan, N. V. Smith, *Phys. Rev. B,* 1985, **31**, 8049.
23. J. A. Wilson, *Solid State Commun*. 1977, **22**, 551.
24. T. E. Kidd, T. Miller, M. Y. Chou, T. C. Chiang, *Phys. Rev. Lett*, 2002, **88**, 226402.
25. N. Giang, Q. Xu, Y. S. Hor, A. J. Williams, S. E. Dutton, H. W. Zandbergen, R. J. Cava, *Phys. Rev. B,* 2010, **82**, 024503.
26. J. Rodríguez-Carvajal, *Comm. Powder Diffr.,* 2001, **26**, 12.
27. E. Morosan, H. W. Zandbergen, B. S. Dennis, J. W. G. Bos, Y. Onose, T. Klimczuk, A. P. Ramirez, N. P. Ong and R. J. Cava, *Nat. Phys.,* 2006, **2**, 544.
28. H. X. Luo, W. W. Xie, J. Tao, I. Pletikosic, T. Valla, G. S. Sahasrabudhe, G. Osterhoudt, E. Sutton, K. S. Burch, E. M. Seibel, J. W. Krizan, Y. M. Zhu and R. J. Cava, *Chem. Mater.,* Under reviewed, 2016.
29. W.L. McMillan, *Phys. Rev.*, 1968, **167(2),** 331.

**Figure Captions**

**Figure 1.** Comparison of the starting material Pb/Ti and Sn/Ti ratios with the EDX analysis of the single crystals grown of the misfit phase $(Pb_{1-x}Sn_xSe)_{1.16}(TiSe_2)_2$ ($0 \leq x \leq 0.6$).

**Figure 2**. XRD pattern showing the (00L) reflections for a selected single crystal of $(Pb_{0.6}Sn_{0.4}Se)_{1.16}(TiSe_2)_2$. Inset, an example of a crystal plate of $(Pb_{0.6}Sn_{0.4}Se)_{1.16}(TiSe_2)_2$.

**Figure 3**. Schematic of the crystal structure of $(Pb_{1-x}Sn_xSe)_{1.16}(TiSe_2)_2$ ($x = 0.4$), (a) along the b-direction; and (b) along the a-direction. The figures highlight the rocksalt double layers and the two 1T-like $TiSe_2$ layers whose interlayering generates the crystal structure of the misfit phase, while also showing that the in-plane matching of the layers in these directions is incommensurate.

**Figure 4. Transport characterization of the normal states and superconducting transitions** (a) The temperature dependence of the resistivity ratio ($\rho/\rho_{300K}$) for polycrystalline $(Pb_{1-x}Sn_xSe_2)_{1.16}(TiSe_2)_2$ ($0.0 \leq x \leq 0.6$), including the low temperature region (2 – 5.5 K); (b) $d\rho/dT$ for $(Pb_{1-x}Sn_xSe_2)_{1.16}(TiSe_2)_2$ ($0.0 \leq x \leq 0.6$) in the low temperature region (2 – 5.5 K), showing the information used to determine the superconducting transition temperature.

**Figure 5. Low temperature specific heat characterization of $(Pb_{0.6}Sn_{0.4}Se)_{1.16}(TiSe_2)_2$.** Temperature dependence of the specific heat $C_p$ of a single crystal of $(Pb_{0.6}Sn_{0.4}Se)_{1.16}(TiSe_2)_2$ measured under magnetic fields of 0 T and 5 T, presented in the form of $C_p/T$ vs $T^2$. The values of γ and β (see text) were obtained by fitting the heat capacity data obtained in the range 2-5 K in the magnetic field of 5 T. The inset shows the electronic specific heat and the equal area construction employed to determine $\Delta C/\gamma T_c$.

**Figure 6. The $T_c$ vs. Sn content x of the superconductor in the $(Pb_{1-x}Sn_xSe)_{1.16}(TiSe_2)_2$ misfit system.** The composition of the trivial to topological transition in rocksalt $Pb_{1-x}Sn_xSe$ is shown by a dashed line.

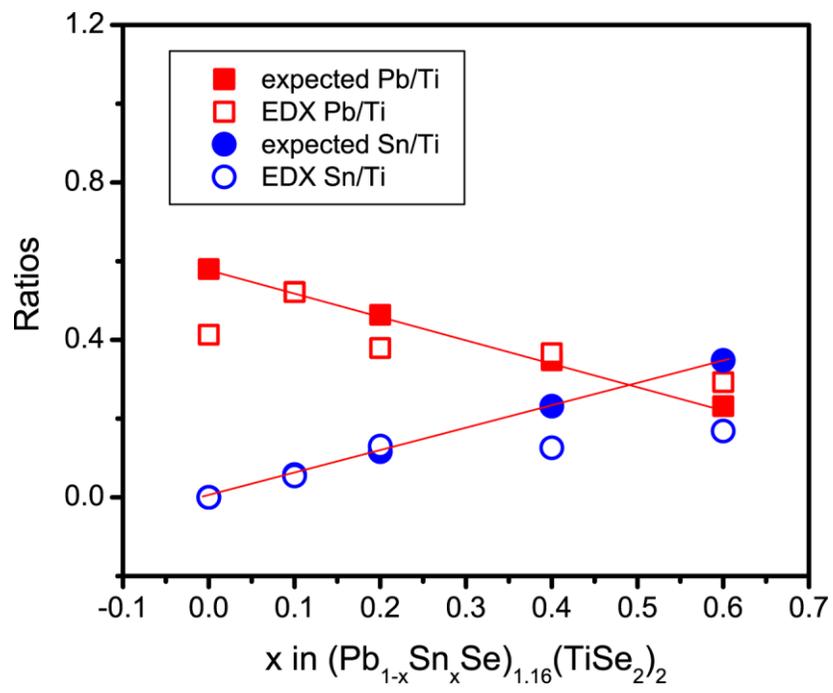

**Figure 1.**

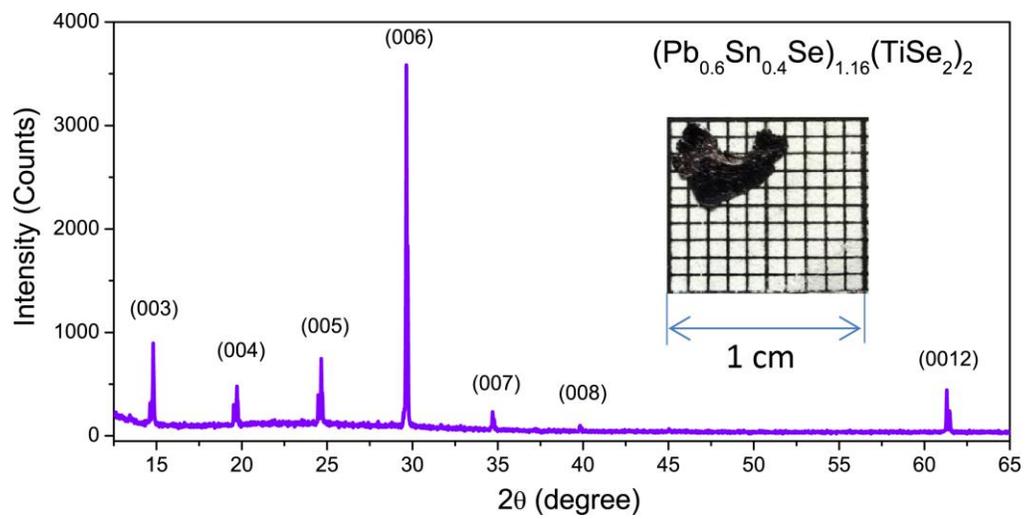

**Figure 2**.

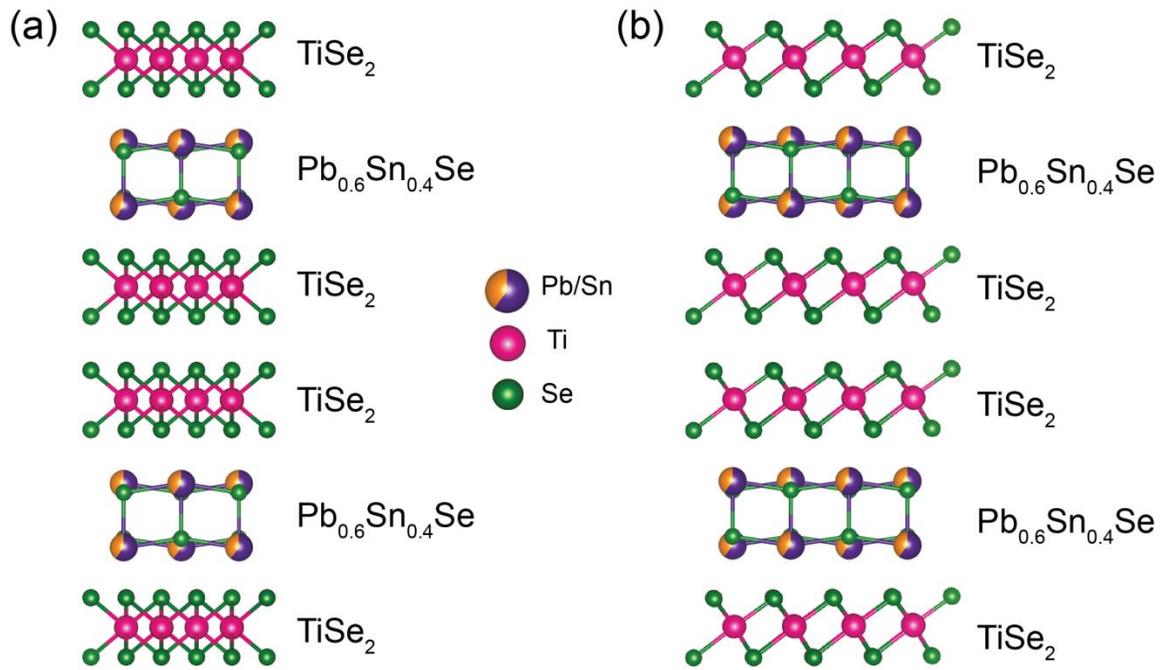

**Figure 3**.

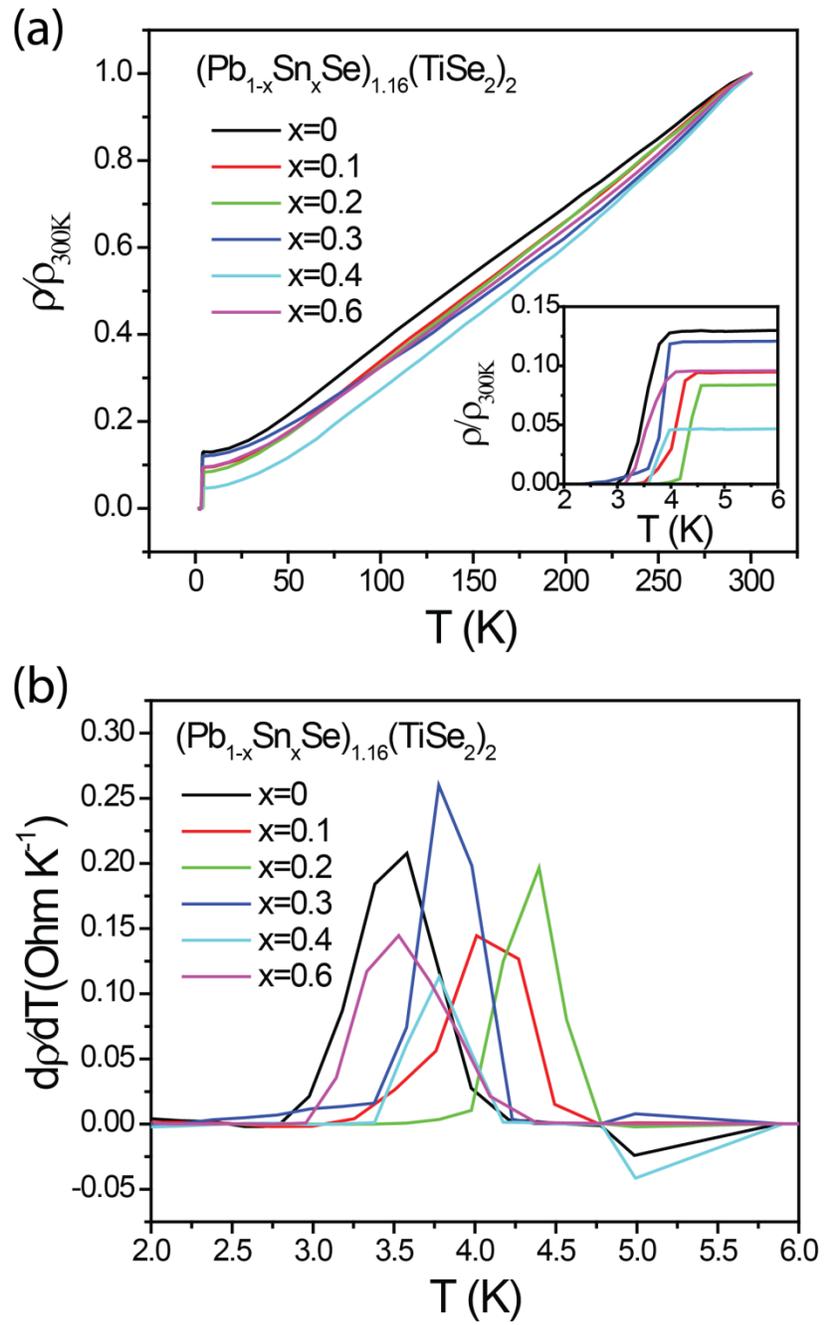

Figure 4.

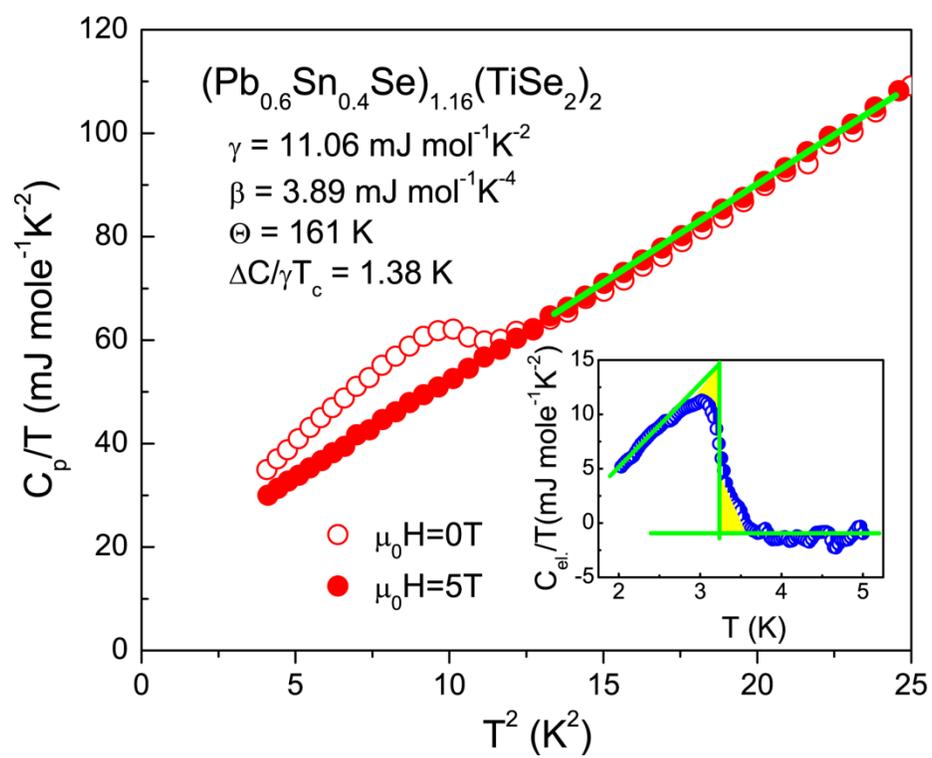

**Figure 5.**

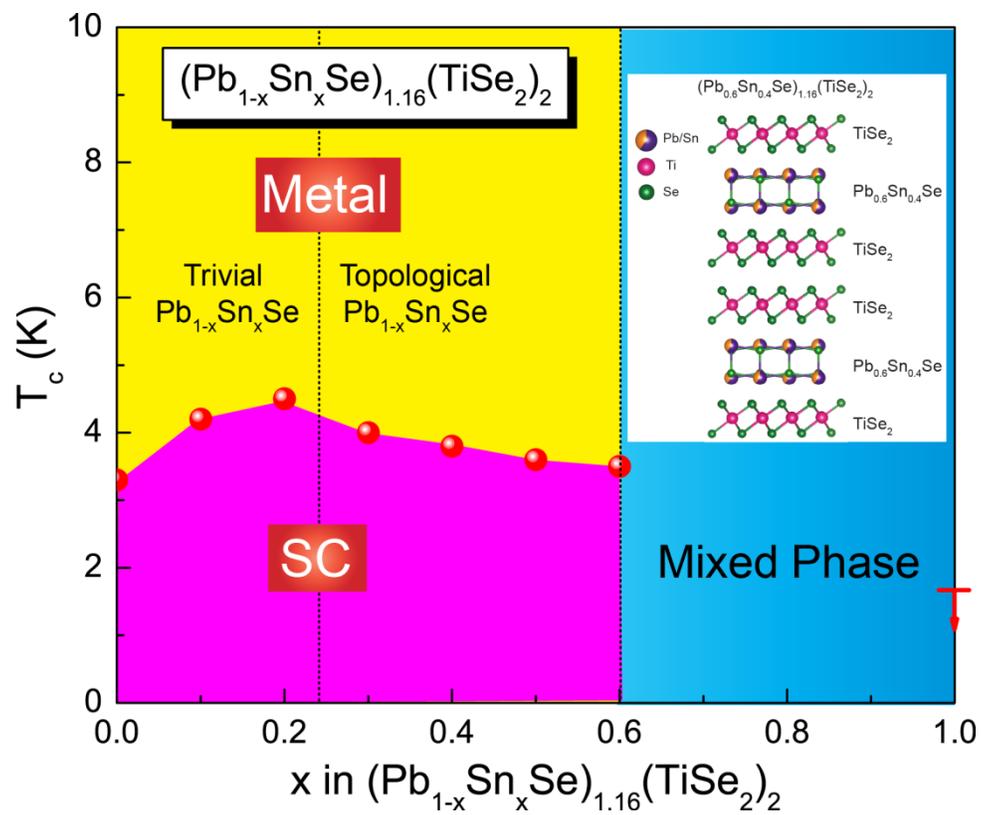

Figure 6.